\documentclass[9pt,twoside]{article}
\pdfoutput=1
\usepackage{epsfig}
\usepackage{calc}
\usepackage{subfigure}
\usepackage{calc}
\usepackage{amssymb}
\usepackage{amstext}
\usepackage{amsmath}
\usepackage{multicol}
\usepackage{times}
\usepackage{apalike}
\usepackage{fancyhdr}
\usepackage{iceis}

\usepackage{ifpdf}
\usepackage{color}
\ifpdf
\else

\fi

\begin{document}
\onecolumn \maketitle \normalsize \vfill 

\section{INTRODUCTION}
\label{sec:Introduction}

Data warehousing and OLAP (On-Line Analytical Processing) technologies~\cite{INM02,KIM02} are now considered mature. They are aimed, for instance, at analyzing the behavior of a customer, product, or company, and may help monitoring one or several activities (commercial or medical pursuits, patent deposits, etc.). More precisely, they help analyzing these activities under the form of numerical data. However, in real life, many decision support fields (customer relationship management, marketing, competition monitoring, medicine...) need to exploit data that are not only numerical or symbolic.
We term such data complex data. Their availability is now very common, especially since the broad development of the World Wide Web.
For example, a medical file is usually constituted of several pieces of data under various forms. A patient's medical history might be recorded as plain text. Various biological exam results might be indicated in many ways. The file could also include radiographies (images) or echographies (video sequences). Successive diagnoses and therapies might be recorded as text or audio documents, etc. Another example could be a collection of web documents concerning a given topic, which would be available under various formats (videos, images, sounds, texts, etc.).


Complex data might be structured or not, and are often located in different and heterogeneous data sources. Browsing these data necessitates an adapted approach to help collect, integrate, structure and eventually analyze them. A data warehousing solution is interesting in this context, though adaptations are obviously necessary to take into account data complexity. Measures might not necessarily be numerical, for instance. Data volumetry and dating are also other arguments in favor of the warehousing approach.
Furthermore, complex data produce different kinds of information that are represented as metadata. These metadata, along with domain-specific knowledge, are essential when processing complex data and play an important role when integrating, managing, or analyzing them. Hence, metadata need to be given even more importance than in classical data warehousing.

The notion of complex data is not straightforward. To clarify this concept, we propose in this paper one definition that presents the different aspects we identify in complex data (Section~\ref{sec:DefinitionOfComplexData}). This definition is, to the best of our knowledge, somewhat complete and pertinent, but its scope could certainly be widened.
We also propose a general architecture framework for warehousing complex data (Section~\ref{sec:ProposedArchitecture}). This model heavily relies on metadata and domain-specific knowledge. It also rests on the XML language that we use for different purposes:
 to store complex data, if necessary;
 to store metadata and knowledge about these complex data;
 and to facilitate communication between the different warehousing processes --- ETL (Extract, Transform, Load) and integration, administration and monitoring, and analysis and usage.
We finally conclude this paper and provide research perspectives (Section~\ref{sec:ConclusionAndPerspectives}).

\section{A DEFINITION OF COMPLEX DATA}
\label{sec:DefinitionOfComplexData}

Many researchers in several communities start to claim they work on complex data. However, this emerging concept of complex data varies a lot, even within a single research community such as the database community. Hence, in a first step, we performed an extensive litterature study to identify all the different sorts of data researchers dealt with. We particularly, but not exclusively, focused on publications and events that explicitely mentionned the terms ``complex data", which particularly emerge in the data mining field~\cite{GAN04}. After compiling all this information, we were able to propose a first definition and concluded that data could be qualified as complex if they were:
(1) \emph{multiformat}, i.e., represented in various formats (databases, texts, images, sounds, videos...);
and/or
(2) \emph{multistructure}, i.e., diversely structured (relational databases, XML document repositories...);
and/or
(3) \emph{multisource}, i.e., originating from several different sources (distributed databases, the Web...);
and/or
(4) \emph{multimodal}, i.e., described through several channels or points of view (radiographies and audio diagnosis of a physician, data expressed in different scales or languages...);
and/or
(5) \emph{multiversion}, i.e., changing in terms of definition or value (temporal databases, periodical surveys...).

\begin{figure}[hbt]
    \centering
    \includegraphics[width=7cm]{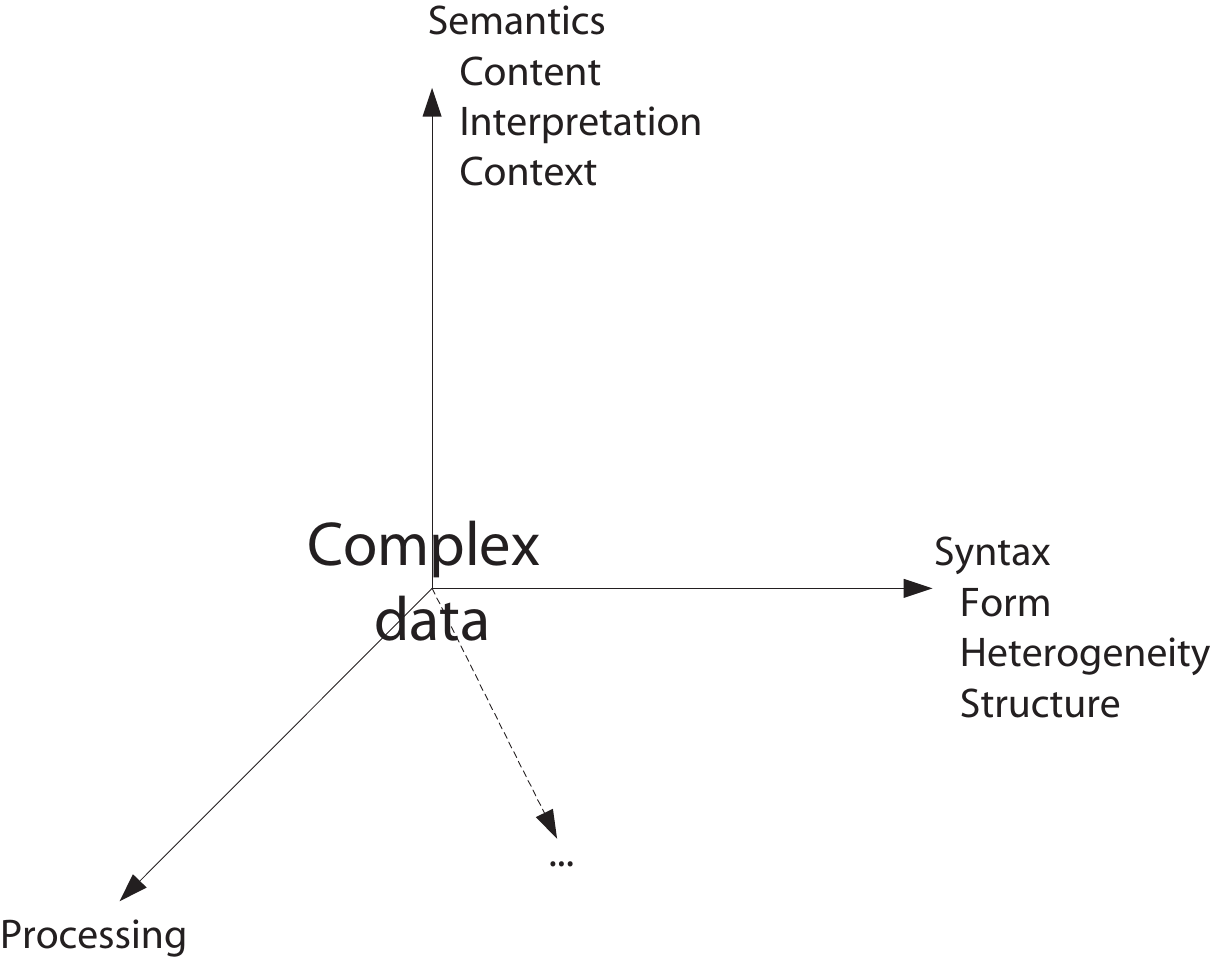}
    \caption{Axes of data complexity}
    \label{fig:complexdata}
\end{figure}

However, it appeared in subsequent meetings with fellow researchers that this first definition was not sufficient to cover the wide variety of complex data. It could indeed be viewed as an axis of complexity, among other axes dealing with data semantics or processing, for instance (Figure~\ref{fig:complexdata}).
Data volumetry could also be such an axis. Though data volume is not an expression of intrinsic complexity if viewed in terms of database tuples, it becomes a complex problem to deal with in statistics or data mining when it is the number of attributes that increases.
In conclusion, we define in this section a framework that helps identifying what we term complex data. However, since this definition cannot be exhaustive, we leave it open to new axes of complexity.

\section{COMPLEX DATA WAREHOUSE ARCHITECTURE FRAMEWORK}
\label{sec:ProposedArchitecture}

In opposition to classical solutions, complex data warehouse architectures may be numerous and very different from one another. However, two approaches seem to emerge.

The first, main family of architectures is data-driven and based on a classical, centralized data warehouse where data are the main focus. XML document warehouses~\cite{XYL01,BAR03,HUM03,NAS04} are a examples of such solutions. They often exploit XML views, which are XML documents generated from whole XML documents and/or parts of XML documents. A data cube is then a set of XML views. 

The second family of architectures includes solutions based on virtual warehousing, which are process-driven and where metadata play a preeminent role. These solutions are based on mediator-wrapper approaches and exploit distributed data sources. These sources' schemas are one of the main information mediators exploit to answer user queries. Data are collected and multidimensionnally modeled (as data cubes, to constitute OLAP analysis contexts) on the fly to answer a given decision support need~\cite{AMM01}.

Whatever the type of architecture, the various processes used in data warehousing always deal with metadata and domain-specific knowledge, in order to achieve a better exploitation and good performances.
Note that complex data are generally represented by descriptors that may either be low-level information (an image's size, an audio file's duration, the speed of a video sequence...) or relate to semantics (relationships between objects in a picture, topic of an audio recording, identification of a character in a video sequence...). Processing the data thus turns out to process their descriptors. Original data are stored, for instance as binary large objects (BLOBs), and can also be exploited to extract information that could enrich their own caracteristics (descriptors and metadata).

\begin{figure*}[hbt]
    \centering
    \includegraphics[width=11.5cm]{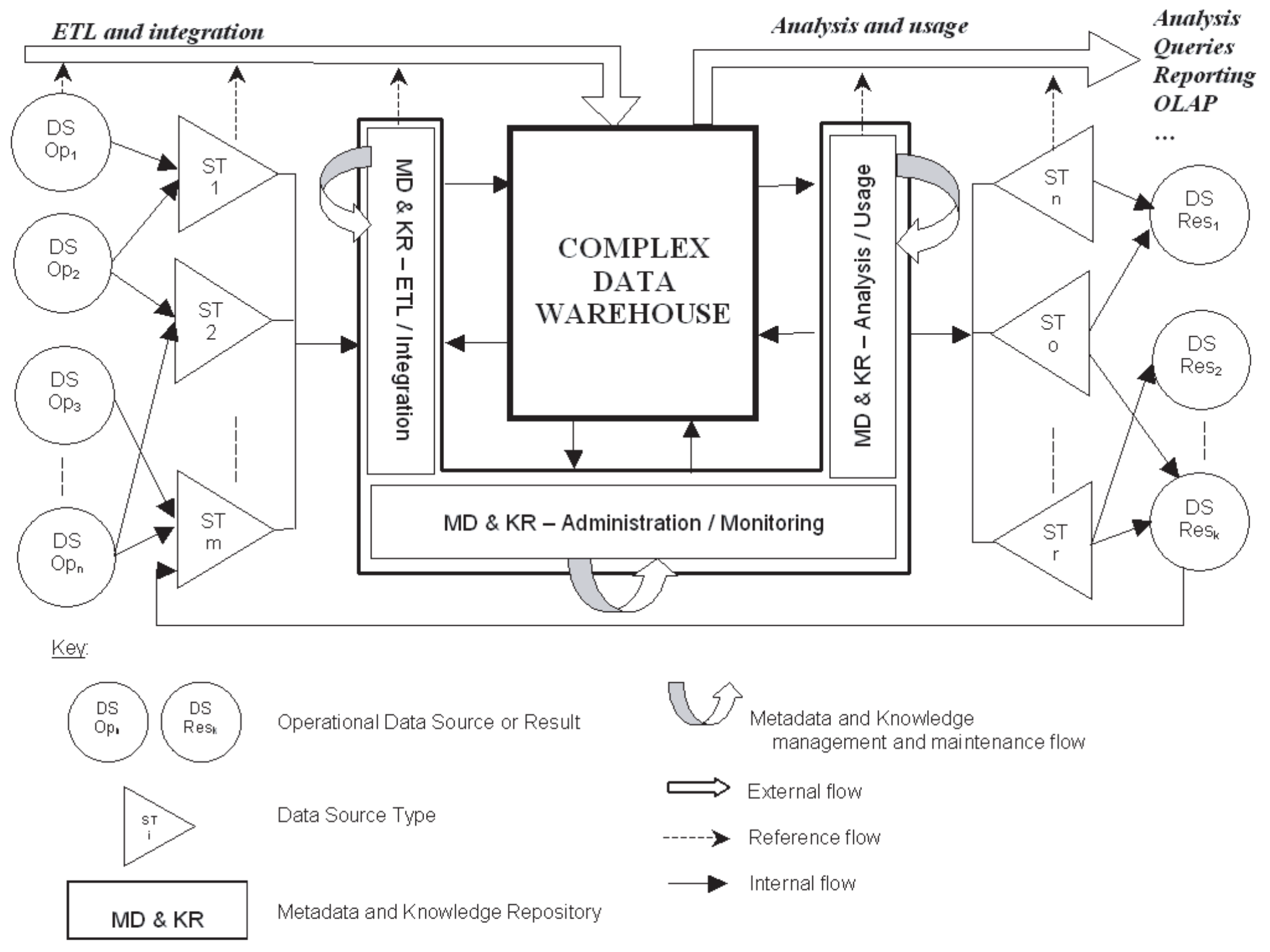}
    \caption{Complex data warehouse architecture framework}
    \label{fig:archi}
\end{figure*}

The architecture framework we propose for complex data warehousing (Figure~\ref{fig:archi}) exploits the XML language.
Using XML indeed facilitates
     the integration of heterogeneous data from various sources into the warehouse;
     the exploitation of metadata and knowledge (namely regarding the application domain) within the warehouse;
     and data modeling and storage.
The presence of metadata and knowledge in the data warehouse is aimed at improving global performance, even if their actual integration is still the subject of several research projects~\cite{MCB01,BAR03,SHA03}.

This architecture framework is essentially made of:
the \textit{data warehouse kernel}, which may be either materialized as an XML warehouse, or virtual (where cubes are computed at run time);
\textit{data sources};
\textit{source type drivers} that notably include mapping specifications between the sources and XML;
and a \textit{metadata and knowledge base layer} that includes three submodules related to three management processes.

The three processes for managing a data warehouse are:
the \textit{ETL and integration} process that feeds the warehouse with source data from operational databases ($DS~Op$) by using drivers that are specific to each source type ($ST$);
the \textit{administration and monitoring} process ($MD \& KR$) that manages metadata and knowledge (the administrator interacts with the data warehouse through this process);
and the \textit{analysis and usage} process that runs user queries, produces reports, builds data cubes, supports OLAP, etc.
Each of these processes exploits and updates the metadata and the knowledge base. There are four types of flows:
the \textit{external flow}, which includes the ETL and integration flow and the exploitation (analysis and usage) flow (the warehouse may thus be considered as a black box);
the \textit{internal flow}, between the warehouse kernel and the metadata and knowledge base layer and between the metadata and knowledge base layer and the source type drivers;
the \textit{metadata and knowledge management and maintenance flow}, which acquires new knowledge and enriches existing knowledge;
and the \textit{reference flow}, which illustrates the fact that the external flow always refers to the metadata and knowledge base layer for integration, ETL, and analysis and usage in general.

Note that analysis results under the form of cubes, reports, queries, or any other intermediary results may constitute new data sources ($DS~Res$) that may be reintegrated into the warehouse.

Though our proposal is only an architecture framework, it helps us formalizing the warehousing process of complex data as a whole. Thus, we are able to identify the issues to be solved. We can also point out the great importance of metadata in managing and analyzing complex data. Furthermore, piloting and synchronizing the data warehouse processes we identify in this framework is a whole problematic in itself. Optimization techniques will be necessary to achieve an efficient management of data and metadata. Communication techniques, presumably based on known protocols, will also be needed to build up efficient data exchange solutions.

\section{CONCLUSION AND PERSPECTIVES}
\label{sec:ConclusionAndPerspectives}

We addressed in this paper the problem of warehousing complex data. We first clarified the concept of complex data by providing a precise, though open, definition of complex data. Then we presented a general architecture framework for warehousing complex data. It heavily relies on metadata and domain-specific knowledge, which we identify as a key element in complex warehousing, and rests on the XML language, which helps storing data, metadata and knowledge, and facilitates communication between the various warehousing processes. This proposal takes into account the two main possible families of architectures for complex data warehousing (namely virtual data warehousing and centralized, XML warehousing). Finally, we rapidly presented the main issues in complex data warehousing, especially regarding data integration, the modeling of complex data cubes, and performance.

This study opens many research perspectives. Up to now, our work mainly focused on the integration of complex data in an ODS. Though we also worked on the muldimensional modeling of complex data, this was our first significant advance into the actual warehousing of complex data. In order to test and refine our hypotheses in the field, we plan to apply our proposals on three different application domains we currently work on (medicine, banking and geography). Such practical applications should help us devise solutions about the many issues regarding metadata management and performance, and experiment both the virtual and XML warehousing solutions.

One of our important perspectives deals with the selection of a representation mode for metadata and domain-specific knowledge. Knowledge related to the application domain is actually an operational information about complex data. It may be considered as metadata. In order to remain in the XML-based, homogeneous environment of our architecture framework, the formalisms that seem best-fitted to represent metadata are XML and RDF (Resource Description Framework) schemas. These tools are appropriate to represent both low-level and semantic descriptors. Furthermore, they are adapted to reasoning for metadata exploitation. The Common Warehouse Metamodel (CWM), an OMG standard for data warehouses~\cite{OMG03}, could also help us managing metadata and knowledge. But can the CWM metamodels integrate the performance factors of a complex data warehouse? Should these metamodels be extended or would it be more interesting to propose new submodels instead? These are largely open questions.

\bibliographystyle{apalike}
\small
\bibliography{complexdw}

\end{document}